\documentclass{article}

\usepackage[preprint,nonatbib]{neurips_2024}

\usepackage[utf8]{inputenc} 
\usepackage[T1]{fontenc}    
\usepackage{hyperref}       
\usepackage{url}            
\usepackage{booktabs}       
\usepackage{amsfonts}       
\usepackage{nicefrac}       
\usepackage{microtype}      
\usepackage{xcolor}         
\usepackage{makecell}
\usepackage{graphics}
\usepackage{subfigure}
\usepackage{graphicx}
\usepackage{biblatex}
\usepackage{float}
\usepackage{amsmath}
\usepackage{multirow}

\usepackage{color}

\title{ZeroPP: Unleashing Exceptional Parallelism Efficiency through Tensor-Parallelism-Free Methodology}

\author{%
  Ding Tang,
  Lijuan Jiang\thanks{Corresponding author.}, 
   Jiecheng Zhou,
   Minxi Jin
   \And
  Hengjie Li,
  Xingcheng Zhang,
  Zhilin Pei\\ \\
  Shanghai Artificial Intelligence Laboratory \\
  \{tangding,jianglijuan,zhoujiecheng,jinminxi,lihengjie,zhangxingcheng,peizhilin\}@pjlab.org.cn
  \And
    Jidong Zhai \\ \\
  Tsinghua University\\
  zhaijidong@tsinghua.edu.cn
}

\begin{document}

\maketitle

\begin{abstract}
Large-scale models rely heavily on 3D parallelism for distributed training, which utilizes tensor parallelism (TP) as the intra-operator parallelism to partition model states across GPUs. However, TP introduces significant communication overheads and complexity in modifying single-GPU code. In this paper, we propose a TP-free distributed framework ZeroPP, which leverages the hybrid of scalable inter-operator pipeline parallelism and
intra-operator fully sharded data parallelism to train models at scale, reducing memory
consumption and enabling high training efficiency.
Through extensive experimentation, we demonstrate that ZeroPP achieves significant performance gains of up to 33\% compared to conventional 3D parallelism while maintaining comparable GPU memory consumption.
\end{abstract}

\section{Introduction} 
Large-scale fundamental models such as BERT~\cite{kenton2019bert}, GPT~\cite{brown2020language}, LLaMA~\cite{touvron2023llama} are flourishing in AI areas by utilizing massive datasets and stacked transformer-based architectures. Striking performance gains have been demonstrated in various areas such as NLP~\cite{kenton2019bert,brown2020language,touvron2023llama}, CV~\cite{dosovitskiy2020image}, etc, with giant model sizes. In recent years, tens to hundreds of billions of model parameters gradually become common in large-scale network training, and the model scale still grows rapidly~\cite{du2021all}. Inevitably, efficient distributed parallel parallelism becomes the crucial fundamental support for large-scale network development, due to the enormous training time and the explosive growth of model sizes far beyond the memory capacity of a single accelerator.

With the increasing demand for distributed training brought about by larger model sizes, various parallel strategies have been proposed, such as data parallelism (DP)~\cite{dean2012large}, pipeline parallelism (PP)~\cite{huang2019gpipe,narayanan2019pipedream}, and tensor parallelism (TP) ~\cite{shoeybi2019megatron}. In addition, concerning memory limitations and training efficiency, hybrid parallel strategies are further utilized to support the training of large-scale models with parameter counts ranging from tens of billions to trillions on hundreds to thousands of GPUs~\cite{smith2022using,xu2020megatron}. In particular, as
the de facto large-scale distributed method, 3D parallelism~\cite{narayanan2021efficient} has been widely used, which combines DP, TP, and PP to achieve high training efficiency by weighing communication, computation, and memory usage.                

TP as an intra-operator parallelism is leveraged by 3D parallelism to distribute model states across GPUs~\cite{2021Colossal}. However,
TP necessitates heavy collective communication for forward and backward computation since the weight tensors are partitioned along the reduction dimension.
Therefore, TP is typically performed across GPUs within a node with high bandwidth~\cite{narayanan2021efficient}. Moreover, implementing TP requires intricate modification to the single GPU code and poses a substantial burden on data scientists and AI practitioners, hindering its widespread adoption and usability.

\begin{figure}[ht]
\begin{center}
\centerline{\includegraphics[width=0.7\linewidth]{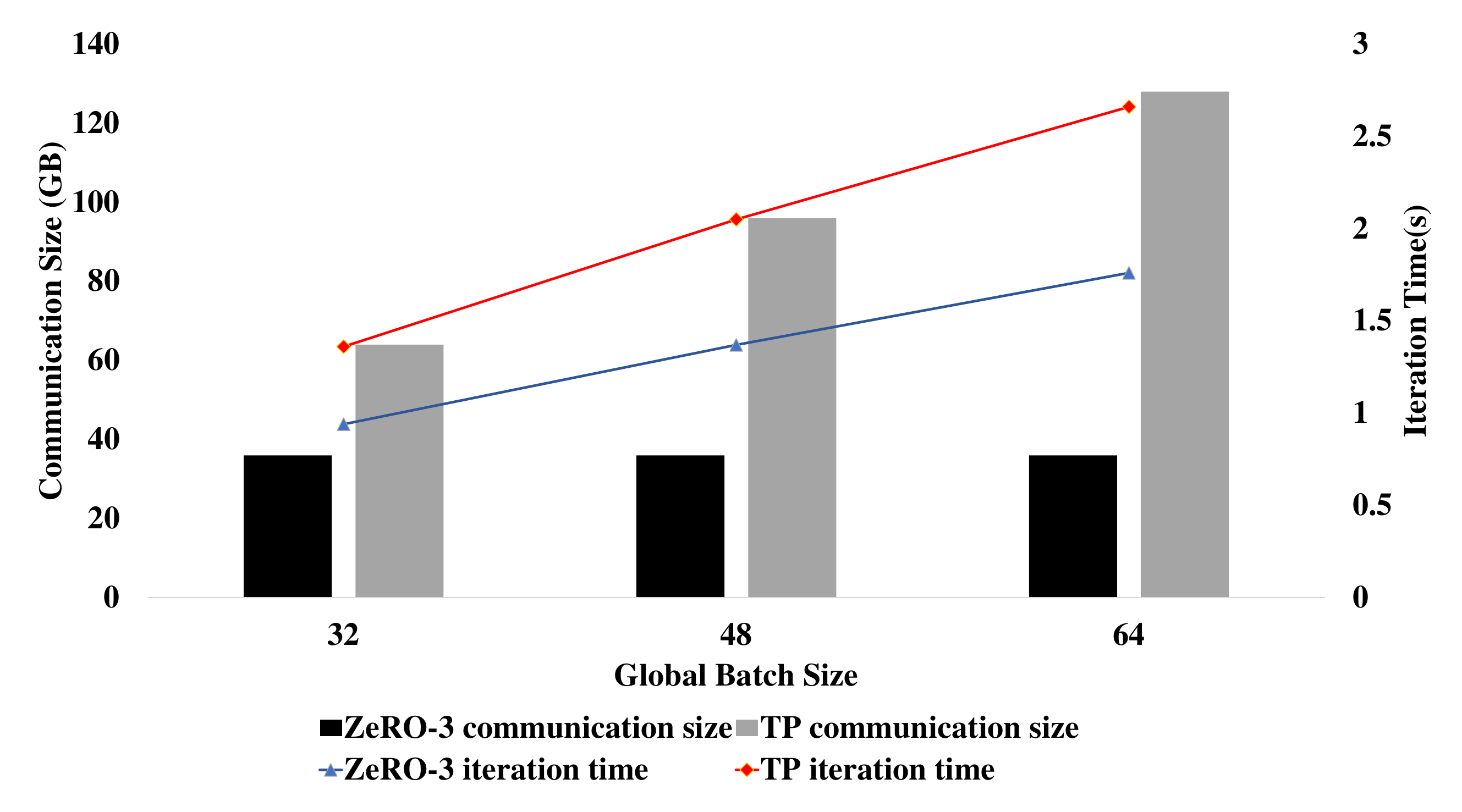}}
\caption{
Communication efficiency comparison between TP and ZeRO-3.
We test the 6.2B GPT model parallelized with the two methods, respectively, across 8 GPUs
with varying global batch sizes, and estimate the per-GPU communication volume for each method. 
Consequently, ZeRO-3 gains superior communication efficiency, and the performance gap becomes more pronounced as the global batch size increases.}
\label{fig:tpzero}
\end{center}
\vskip -0.2in
\end{figure}

ZeRO is another widely used intra-operator parallelism method, a refined variant of data parallelism that does not necessitate modifications to the model implementations~\cite{rajbhandari2020zero}. ZeRO typically includes multiple stages. ZeRO stage 1 (ZeRO-1) solely partitions the optimizer states, exhibiting consistent communication volume compared to vanilla DP. ZeRO stage 3 (ZeRO-3) partitions the optimizer states, gradients, and parameters, and requires extra parameter gathering before each forward and backward computation. 
But ZeRO-3  remains more communication-efficient than TP. We demonstrate that through a simple experiment in Figure \ref{fig:tpzero}, and provide a more detailed theoretical analysis in Section \ref{sec:4}.

Considering that ZeRO-3 partitions the full set of model states and achieves superior communication efficiency as mentioned above, the hybrid parallelization of ZeRO-3 and PP is expected to outperform the hybrid parallelization of TP and PP. However, the combination of ZeRO-3 and PP as a hybrid parallelism method entails significant repetitive parameter gathering and gradient synchronization due to frequent micro-batch computation switches in PP. The high-frequency communication results in poor efficiency.
Lamy-Poirier et al.~\cite{lamy2023breadth} propose a Breadth-first PP (BF-PP) method which combines ZeRO-3 efficiently but only targets scenarios with small batch sizes.
More specifically, the GPU memory requirements of BF-PP increase infinitely with the growth of batch sizes and model scales. The drawback in memory consumption of BF-PP makes it unpractical in large-scale training since most models are trained at large batch sizes and model scales currently. Besides, as with most PP methods, BF-PP suffers from pipeline bubbles which hurts the training efficiency to a great extent.

In this paper, we propose a TP-free hybrid parallelism method ZeroPP, which combines a scalable PP method and ZeRO-3 to achieve balanced computation, communication, and memory utilization. By employing a blockwise task-interleaved PP schedule, we reduce parameter gathering during computation and communication between PP and ZeRO-3. Memory cost is kept low through unit scheduling, and an efficient recomputation scheme is designed to optimize activation memory usage further. In addition, ZeroPP utilizes a near-zero bubble PP schedule by leveraging gradient computation decoupling. Our contributions are as follows:
\begin{itemize}
    \item We present an efficient PP method that seamlessly integrates with ZeRO-3, ensuring efficient computation, communication, and memory utilization. By carefully decoupling gradient computations and reusing memory through unit scheduling, our approach reduces the memory footprint and promotes the utilization of computational resources outstandingly.
    
    \item We propose a hybrid parallel framework ZeroPP, which combines the newly proposed PP, ZeRO-3, and vanilla DP and is freed from the sophisticated and heavy communication of TP. ZeroPP leverages the hybrid of scalable inter-operator pipeline parallelism and intra-operator fully sharded data parallelism to train models at scale, reducing memory consumption and enabling high training efficiency. 

    \item We involve an effective recomputation method for ZeroPP, which further reduces memory consumption while introducing a marginal increase in computational overhead.
    
    \item We conduct a series of experiments on transformer models of various sizes.
    The results demonstrate that ZeroPP achieves significant performance gains of up to 33\% compared to conventional 3D parallelism while maintaining comparable GPU memory consumption.
    
\end{itemize}

\section{Related Works}

\subsection{Data, Model and 3D parallelism}
DP~\cite{dean2012large} is popular to accelerate neural network training, where each model replica works independently but synchronizes gradients through all-reduce to update the parameters. However, state-of-the-art large-scale networks rely on vast model sizes and training data, which may make DP alone unusable due to memory constraints. 

Model parallelism is introduced to mitigate memory limitations. TP and PP are two typical model parallelism methods. TP~\cite{shoeybi2019megatron,wang2022tesseract,xu2023efficient,wang20212,bian2021maximizing} splits parameters into shards along a specific dimension and each device only holds a shard of each parameter while not affecting the correctness of the computation graph. However, since heavy collective communication overhead is required when the hidden dimension is sharded, TP is often restricted to a single node. PP~\cite{huang2019gpipe,narayanan2019pipedream,narayanan2021efficient,lamy2023breadth} partitions the model into stages, each of which is composed of contiguous network layers and mapped onto a device. PP is widely studied and employed on account of the lightweight communication overhead.

3D parallelism~\cite{shoeybi2019megatron}, integrating DP, PP, and TP, emerges as a compelling solution that attains commendable throughput and scalability. Notably, the technique has proven its efficacy in training diverse, expansive large language models~\cite{black2022gpt,smith2022using}. Nevertheless, the popular 3D parallelism faces substantial obstacles brought by TP as discussed above, i.e. huge complexity to refactor code and a considerable amount of communication overhead.

 
\subsection{ZeRO}

ZeRO~\cite{rajbhandari2020zero} presents a memory-optimized approach to data parallel training by partitioning model states, encompassing parameters, gradients, and optimizer states, across multiple GPUs while selectively retrieving the required model states solely for computing specific layers.

ZeRO entails three distinct stages. ZeRO stage 1 (ZeRO-1) partitions optimizer states across GPUs. In ZeRO stage 2 (ZeRO-2), optimizer states and gradients undergo partition.
Both ZeRO-1 and ZeRO-2 effectively reduce memory consumption while maintaining the same communication volume compared to DP.
 Advancing further, 
ZeRO stage 3 (ZeRO-3) partitions all three constituents of the model states, namely parameters, gradients, and optimizer states. 
ZeRO-3 stands out due to its remarkable memory efficiency but suffers from parameter gathering in each forward and backward computation. Substantial extra communication overheads are introduced.

 To excavate more memory capability, subsequent works such as ZeRO-Infinity~\cite{rajbhandari2021zero} and ZeRO++~\cite{wang2023zero++} have been proposed. However, these approaches rely on CPU offloading or quantization methods, which can introduce performance or accuracy losses.

In this paper, unless otherwise specified, ZeRO refers to ZeRO Stage 3.

\subsection{Gradient Calculation Decoupling in Pipeline Parallelism}
\label{sec:pp}
With the progression of distributed training, numerous PP strategies have been proposed~\cite{huang2019gpipe,narayanan2019pipedream,narayanan2021efficient,lamy2023breadth}. However, a predominant issue encountered by most of these approaches is the presence of pipeline bubbles. One effective solution to mitigate pipeline bubbles involves decoupling the gradient calculations for weights and inputs~\cite {oh2022out,qi2023zero}. ~\footnote{``Parameter'' and ``weight'' are interchangeable in the article.}

In the backward propagation of model training, the loss is backpropagated according to the chain rule of differentiation to calculate input and weight gradients. However, the computation of gradients for inputs and weights can be decoupled. Taking a basic linear layer as an example, the gradients for inputs are obtained by multiplying the gradients of activation values by weights, while the gradients for weights are obtained by multiplying the gradients of activation values by inputs. After decoupling gradient computations, we can first backpropagate the input gradients to maintain data dependencies and then compute the weight gradients to update local parameters where needed in PP.

\section{Method}
\label{sec:3}


\subsection{ZeRO-Compatible Pipeline Parallelism}
\label{subsec:3.1}

\begin{figure}[ht]
\begin{center}
\centerline{\includegraphics[width=1.0\linewidth]{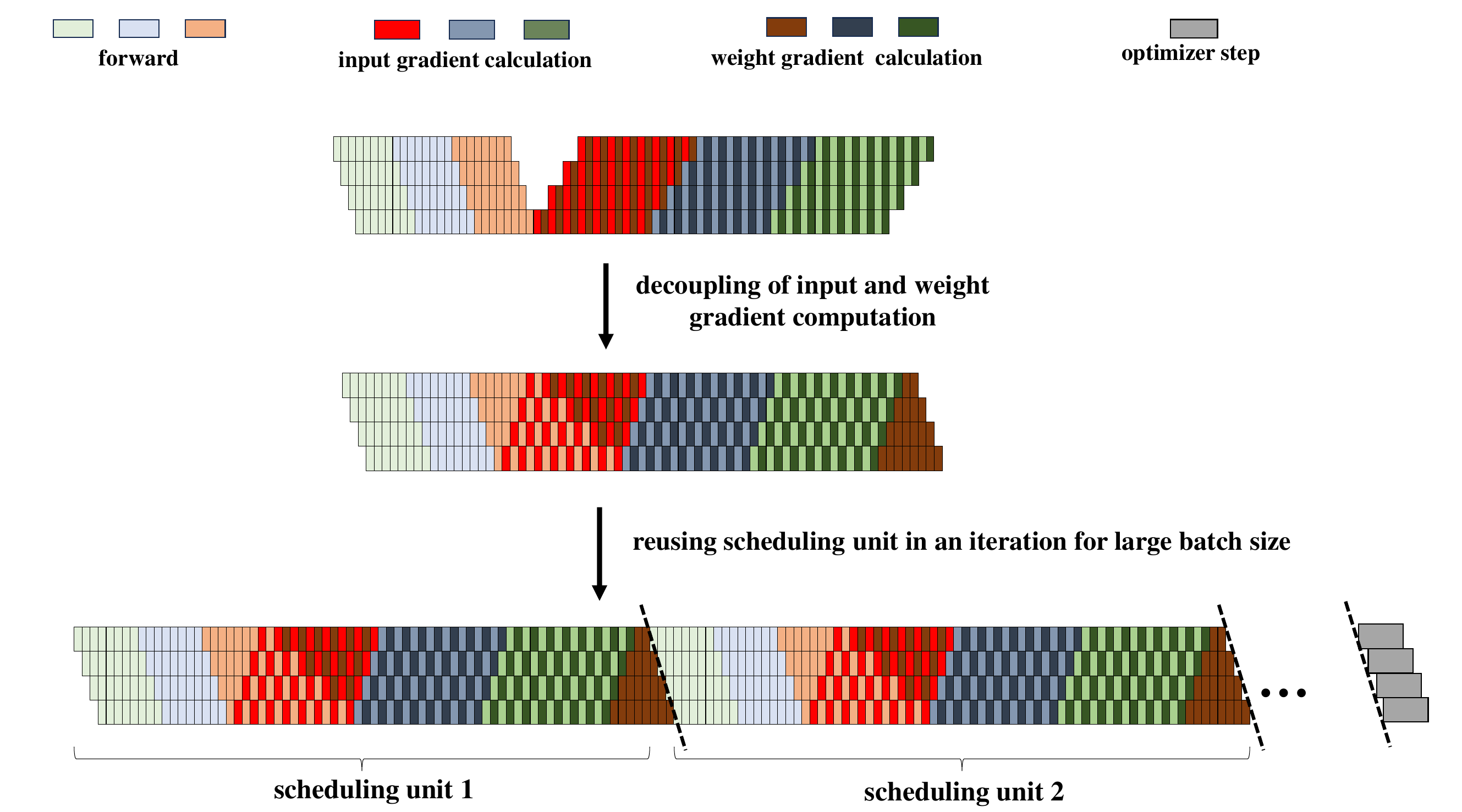}}
\caption{Illustration of the proposed ZeRO-compatible PP schedules. In the three PP schedules presented from top to bottom, effective solutions are utilized pertaining to 
communication, pipeline bubbles, and memory usage at larger batch sizes in the integration of PP and ZeRO. In particular, the micro-batches are divided into multiple groups, and the computations corresponding to each micro-batch group are regarded as a scheduling unit.}
\label{fig:pp}
\end{center}
\vskip -0.2in
\end{figure}

In this section, the designed ZeRO-compatible PP schedules are introduced. As depicted in Figure~\ref{fig:pp}, we illustrate three schedules in a top-to-bottom arrangement, each building upon the improvements of its predecessor. 

\begin{figure}[ht]
\begin{center}
\centerline{\includegraphics[width=1.0\linewidth]{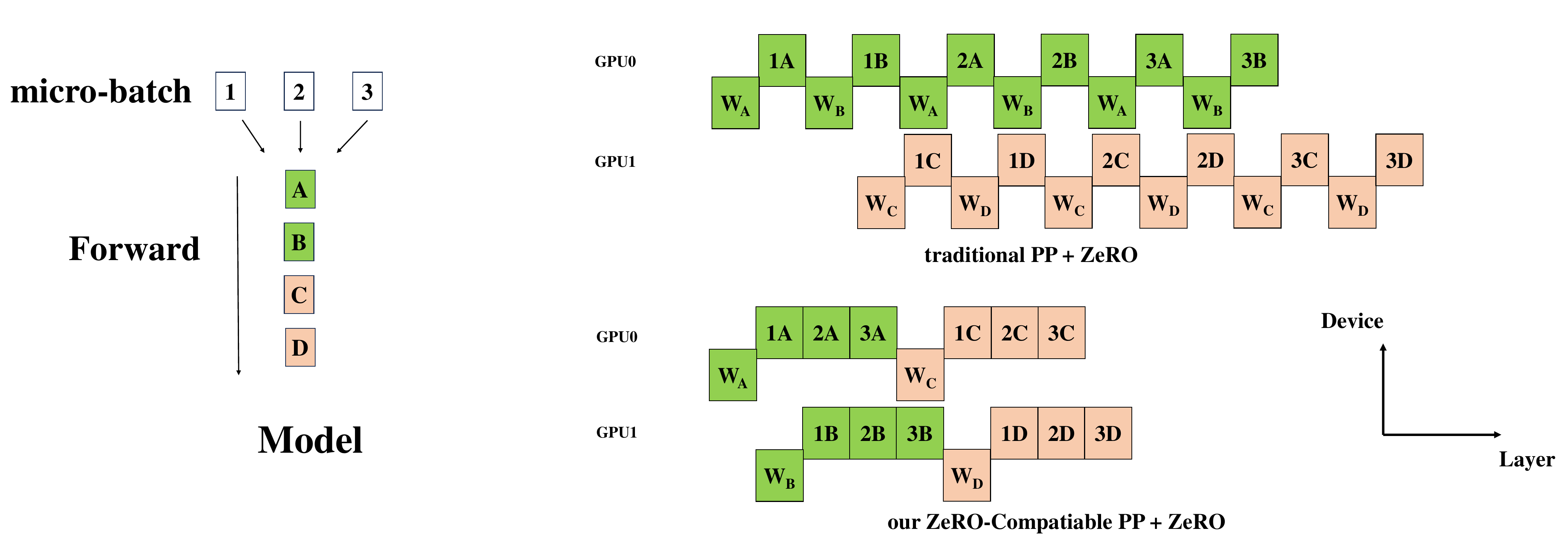}}
\caption{Illustration on forward pass communications in the combination of PP and ZeRO. Digits from 1 to 3 are the indices of micro-batches, and letters from A to D are the indices of layers. $W_i$ represents the parameter gathering communication of the $i$-th layer.}
\label{fig:ivsni}
\end{center}
\vskip -0.2in
\end{figure}

We first elaborate on the repetitive parameter gatherings in the combination of conventional PP strategies with ZeRO. 
To enhance comprehension, we depict a concrete example in Figure \ref{fig:ivsni}, showcasing the forward pass of a 4-layer model with 3 micro-batches distributed across 2 GPUs. As observed, in a conventional PP schedule combined with ZeRO, parameter gathering is required before each micro-batch (and likewise gradient reduction in the backward pass), resulting in significant communication overhead and decreased GPU utilization.

A ZeRO-compatible PP schedule shown as the first schedule of Figure~\ref{fig:pp} partitions the model into stages which are multiples of the pipeline parallelism degree and mapped to the devices with a looping placement. The micro-batches of each stage are scheduled consecutively to reuse the gathered parameters in the stage computation of different micro-batches, which reduces parameter gathering times largely as shown in Figure \ref{fig:ivsni}. Similar PP strategy have been proposed~\cite{lamy2023breadth}. However, we observe that such a PP strategy brings about two issues. Firstly, when the batch sizes
are small, pipeline bubbles cost a large proportion of training time, resulting in decreased GPU utilization. Secondly, when the batch sizes
are large,
the occupied memory by activation values cannot be released in a timely manner, leading to out-of-memory (OOM) errors.
Therefore, 
we make two further improvements to enhance its generality.

\subsubsection{Near-Zero Bubble Pipeline Parallelism}\label{nzbpp}
Pipeline bubbles are bunches of device idling time due to the maintenance of data dependencies during forward and backward propagation, which cause performance degradation and restrict the use of pipeline parallelism. We leverage the separation of gradient calculation of inputs and weights for a delicate pipeline schedule~\cite{oh2022out,qi2023zero}, aiming to achieve near-zero bubbles.

Initially, each device is mapped with multiple stages using the looping placement~\cite{narayanan2021efficient}.
Besides, we categorize the stage computation into four types, i.e.
forward pass, input gradient computation, weight gradient computation, and optimizer step. 
Since input gradients are transferred among devices to maintain the data dependency of network layers, we schedule input gradient computations as early as possible to reduce the device idling time. Besides, we insert weight gradients whose data dependency is satisfied into pipeline bubbles to further improve efficiency.

More specifically, the optimized pipeline schedule includes a forward process, a forward-backward-interleaved process, and a backward process as shown in the second schedule of Figure~\ref{fig:ivsni}. In the forward process, micro-batches of forward passes are breadth-firstly scheduled by stages on each device. The forward-backward-interleaved process consists of an interleaved schedule on the calculation tasks of forward, input gradient, and weight gradient corresponding to the last mapped stage in the looping placement. The forward passes and input gradient computations are scheduled with higher priority, while weight gradient computations are used for filling in the pipeline bubbles.
In the backward process, the micro-batch calculations of input and weight gradient are breadth-firstly scheduled by stages, following which the left weight gradient calculations of the last stage are scheduled. Once the number of micro-batches and the pipeline parallelism degree satisfy a certain condition as analyzed in Section~\ref{sec:4}, near-zero pipeline bubbles could be achieved.
The high communication efficiency in the combination of PP and ZeRO is ensured with the blockwise schedule where micro-batches of computations are breadth-firstly scheduled by stages.


\subsubsection{Reusing Allocated Memory of Scheduling Units}

In order to reduce activation memory consumption, we utilize a looping scheduling scheme. The micro-batches are divided evenly into multiple groups. The stage computations are scheduled by the micro-batch groups. We call the stage computations corresponding to each micro-batch group a scheduling unit.
In each scheduling unit, the near-zero bubble pipeline schedule is performed. In addition, the size of the scheduling unit is set adaptively according to the batch size.

The memory allocated by each scheduling unit will be promptly released before the next scheduling unit begins. Hence, the maximum allocated memory is identical to the allocated memory of the scheduling unit. By adaptively setting the size of the scheduling unit according to batch sizes, ZeroPP achieves a satisfying trade-off between memory consumption and communication.

\subsection{TP-Free Hybrid Parallelism}

In this section, we will describe how the proposed PP strategy is integrated with ZeRO and vanilla DP to achieve an efficient hybrid parallelism method for large-scale distributed training.

\subsubsection{ZeRO-3 + PP + DP}
Models are generally trained on large-scale clusters, where the bandwidth between nodes is comparatively lower than that within a node.
Therefore, in ZeroPP, we combine ZeRO-3, PP, and vanilla DP as a hybrid parallelism method to train models at scale in which ZeRO-3 is performed intra-node and DP is performed inter-node.
More concretely, each node stores a complete model state copy that is partitioned evenly inside a node.
In each scheduling unit, two all-gather operations and one reduce-scatter operation for each stage to gather parameters and synchronize gradients, respectively, are performed within a node. We thus avoid parameter gathering between nodes and simply perform one inter-node all-reduce gradient synchronization in the last scheduling unit.

\subsubsection{ZeRO-3 + PP + ZeRO-1}
To further reduce memory usage,
we substitute vanilla DP with ZeRO-1 in the hybrid parallelism method discussed above.
Initially, each node stores a parameter and gradient copy that is partitioned inside a node. The optimizer states are partitioned across all GPUs.
After the gradient computations of the last scheduling unit in each iteration, we perform gradient reduce-scatter across nodes, update the parameters corresponding to the partitioned optimizer states, and perform an all-gather operation to assemble the parameter copy across nodes. In this way, we further partition the optimizer states across the nodes, while maintaining no extra communication overhead.

\subsection{Activation Recomputation}
Recomputation~\cite{chen2016training} is an essential technique for reducing activation memory usage during the training of large-scale models. The technique becomes particularly crucial as the input sequence length of models continues to increase, and the memory consumption associated with activation values grows accordingly. For those PP methods~\cite {oh2022out,qi2023zero} that achieve improvement in pipeline bubbles by scheduling the calculations of input gradients and weight gradients separately, the incorporation of activation recomputation poses a non-trivial challenge. Specifically, when a layerwise recomputation strategy is adopted, both the input and weight gradient computation depend on the activation recomputation. Hence, extra unacceptable recomputation overhead might be introduced due to the separation of gradient computations, thereby limiting further scalability. 

As depicted in Figure \ref{fig:pp}, our proposed PP approach exhibits continuity in the input and weight gradient computation excluding the computations in the forward-backward-interleaved scheduling process.
The continuity requires a single activation recomputation for consecutive input and weight gradient computations without extra overhead. Therefore, building upon the proposed PP approach, we further propose a simple and efficient recomputation scheme in which we perform recomputation for all mapped stages except the last stage in the interleaved scheduling process on each device. The recomputation method can either be full~\cite{chen2016training} or selective~\cite{korthikanti2023reducing}.

\section{Analysis}
\label{sec:4}

In this section, a comprehensive theoretical analysis is conducted on ZeroPP. We will focus on two aspects: comparisons between ZeroPP and conventional 3D parallelism, 
emphasizing the advantages of ZeroPP utilizing the intra-node parallelism ZeRO, 
and comparisons among the PP+DP approaches where intra-node parallelism is not required. We use symbolic representations listed in Table~\ref{tab:symbolics} for ease of description.

\begin{table}[ht]
\caption{The table of symbols used in the paper.}
\label{tab:symbolics}
\vskip 0.15in
\begin{center}
\begin{small}
\begin{tabular}{ll}
\toprule
Symbols & Meanings \\
\midrule
s       &   Sequence length    \\
h       &   Hidden size   \\
L       &   Number of total transformer layers    \\
$M_{w}$  &    Weight memory of a transformer layer    \\
$M_{a}$  &    Activation memory of a transformer layer   \\
V       &   Number of stages on each device    \\
B       &   Number of micro-batches on a single GPU    \\
U       &   Number of micro-batches within a scheduling unit    \\
b       &   Micro-batch size    \\
P       &   PP size    \\
D       &   DP size    \\
\bottomrule
\end{tabular}
\end{small}
\end{center}
\vskip -0.1in
\end{table}

\subsection{Comparison with Conventional 3D Parallelism}

We elaborate with the core transformer block~\cite{vaswani2017attention} with a hidden size of $h$. Besides, 
each device deals with $B$ micro-batches and each micro-batch contains $b$ samples whose sequence length is $s$ in an iteration.
In terms of 3D parallelism,
TP requires 
four all-reduce operations during the forward pass and backward pass, resulting in a communication volume of $8Bbsh$ in a transformer block~\cite{korthikanti2023reducing}.

In ZeroPP, we assume each scheduling unit has the number of micro-batches $U$.
Thus, $\frac{B}{U}$ scheduling units are processed in each iteration. Within each scheduling unit, two all-gather operations are performed for parameter gathering, and one reduce-scatter operation is performed for gradient synchronization, resulting in a communication volume three times that of the parameter size. Considering that the parameter size of a single typical transformer block is around $12h^2$ (we neglect the item with small orders), we can calculate the total communication volume in one iteration as $\frac{36Bh^2}{U}$. Therefore, as long as the condition $2sUb > 9h$ is satisfied, the communication volume of ZeroPP introduced by intra-operator parallelism is smaller than that of conventional 3D parallelism. Current models prefer to have larger sequence lengths and are trained with larger batch sizes, which makes the condition quite easy to meet in practice and enables ZeroPP a competitive hybrid distributed alternative.

In addition,
ZeroPP offers multiple other advantages compared with conventional 3D parallelism. Firstly, the communication in TP is challenging to overlap with computation due to the strong intra-block data dependency, whereas the communication in ZeroPP can be easily overlapped by layerwise data prefetching as mentioned in Section~\ref{sec:3}. Secondly, ZeroPP employs a near-zero bubble PP method, which effectively improves device utilization. Thirdly, ZeroPP presents more communication overlapping opportunities for gradient synchronization in vanilla DP since the stages are scheduled breadth-firstly by micro-batches and can start gradient synchronization earlier compared to conventional PP strategies~\cite{lamy2023breadth}.

\subsection{Comparison among Various PP Methods}
ZeroPP mitigates pipeline bubbles by early scheduling the calculation of input gradients which reduces the waiting time of devices due to data dependency in the backward propagation. Besides, inserting the calculation of weight gradients that are ready in the pipeline bubbles, ZeroPP could achieve near-zero bubbles in the whole training process. 
However, near-zero bubbles are achieved when enough active micro-batches are ensured concerning data dependency in the forward and backward propagation. The minimum number of active micro-batches to achieve near-zero bubbles is $2P-1$. Therefore, pipeline bubbles related to the number of micro-batches $U$ in a scheduling unit are described as

$$
\#Pipeline\,Bubbles = \begin{cases}\approx 0, \quad U \ge 2P-1 \\ \frac{B(2P-1-U)}{U}, \quad U < 2P-1 \end{cases}.
$$
When the number of active micro-batches is less than $2P-1$, pipeline bubbles could still be relieved due to the split of gradient computation and the early schedule of input gradient computation.

The activation memory could be fixed to a maximum memory by a scheduling unit. For the sake of simplicity, we have omitted the analysis of gradients as it follows a similar pattern to that of weight analysis. Therefore, the memory consumption of ZeroPP is described as,

$$
   Max\,Allocation\,Memory = \underbrace{LM_w/(PD)+LM_w/(PV)}_{weight\_mem} + \underbrace{MIN(B, U)LM_a/P}_{activation\_mem}.
$$

We also compare ZeroPP with state-of-the-art PP methods combined with vanilla DP. As listed in Table~\ref{tab:ppcompare}, ZeroPP achieves superior performance in pipeline bubbles and weight memory usage. To further integrate recomputation, the activation memory is reduced largely at a small cost.

\begin{table*}[ht]
\caption{Comparison of state-of-the-art PP methods.}
\label{tab:ppcompare}
\vskip 0.15in
\begin{center}
\begin{small}
\begin{tabular}{lccc}
\toprule
Pipeline Methods & Bubble Ratio & Weight Memory & Activation Memory    \\
\midrule
Gpipe  &  $(P-1)/B \quad \uparrow$ & $LM_w/P \uparrow$ & $\quad \quad  \quad BLM_a/P  \quad \quad \quad \quad \downarrow$   \\
1F1B &  $(P-1)/B \quad \uparrow$ & $LM_w/P \uparrow$ & $\quad\quad\quad\quad LM_a \quad\quad \quad \quad \quad \uparrow\uparrow$ \\
Interleaved 1F1B & $(P-1)/(VB) \uparrow\uparrow$ & $LM_w/P \uparrow$ & $LM_a(1+(P-1)/(VP)) \uparrow $                            \\
ZeroPP   & $\quad$ Near Zero $\quad  \uparrow\uparrow\uparrow$ & $\quad  LM_w(1/(PD)+1/(PV)) \quad  \uparrow\uparrow\uparrow $ & $LM_a(2-1/P) \uparrow$  \\
ZeroPP + Recomp   &$ \quad$ Near Zero $\quad  \uparrow\uparrow\uparrow$ & $\quad  LM_w(1/(PD)+1/(PV))  \quad  \uparrow\uparrow\uparrow$ & $LM_a(2/V-1/(VP)) \uparrow\uparrow\uparrow$ \\
\bottomrule
\end{tabular}
\end{small}
\end{center}
\vskip -0.1in
\end{table*}

\section{Evaluation}
\label{sec:5}

We conduct a series of experiments to validate our main claim that ZeroPP achieves high training efficiency compared to 3D parallelism while maintaining comparable GPU memory consumption. We implement our codebase using PyTorch~\cite{paszke2019pytorch} and conduct our experiments using up to 64 NVIDIA A800-SXM4-80GB GPUs distributed across 8 nodes that are interconnected by an InfiniBand network. The details of GPT models we use are shown in Table \ref{table:models}.

We compare our method with 3D parallelism based on Megatron~\cite{shoeybi2019megatron}. Besides, the 3D parallelism methods incorporated with two state-of-art PP methods are tested, respectively, including the one utilizing the mainstream interleaved 1F1B method~\cite{shoeybi2019megatron} and the other employing the latest ZB-V method~\cite{qi2023zero}. In terms of DP, both our method and 3D parallelism employ the Pytorch DDP~\cite{ddp}.

\begin{table}[h]
\caption{Details of the models}
\label{table:models}
\begin{center}
\begin{small}
\begin{sc}
\begin{tabular}{ccccc}
\toprule
\thead{Model\\Size} & \thead{Num\\Layers} & \thead{Num\\Heads} & \thead{Hidden\\Size} & \thead{Seq\\Length} \\ 
\midrule
14.6B    & 48 & 40 & 5120 & 1024 \\
28.3B    & 64 & 48 & 6144 & 2048 \\
\bottomrule
\end{tabular}
\end{sc}
\end{small}
\end{center}
\end{table}

\subsection{Comparison to 3D parallelism with Fixed Parallelism Degree}

\begin{table}[h]
\caption{Experimental results of ZeroPP versus 3D parallelism under various settings.}
\label{table:main_exp}
\begin{center}
\begin{small}
\begin{tabular}{|c|c|c|c|c|c|} \hline  

\multirow{6}{*}{\shortstack[c]{Setup}}      & Model       &\multicolumn{2}{|c|}{14.6B}  & \multicolumn{2}{|c|}{28.3B}       \\ \hline  
           & DP Size        &\multicolumn{2}{|c|}{1}  & \multicolumn{2}{|c|}{2}    \\  \cline{2-6} 
           & TP/ZeRO-3 Size        &\multicolumn{2}{|c|}{8}  & \multicolumn{2}{|c|}{8}    \\   \cline{2-6}
           & PP Size        &\multicolumn{2}{|c|}{4}  & \multicolumn{2}{|c|}{4}    \\   \cline{2-6}
           & Global Batch Size        &\multicolumn{2}{|c|}{192}  & \multicolumn{2}{|c|}{256}    \\    \hline  
\multirow{5}{*}{\shortstack[c]{Performance}} & Methods &Memory&Iteration Time&Memory&Iteration Time \\ \cline{2-6} 
                           & 3D Parallelism       &10.5G&7.5s&21.6G&12.1s \\ \cline{2-6} 
                           & 3D Parallelism + SP       &9.6G &9.1s&18.9G&11.9s \\ \cline{2-6} 
                           & ZB-V + TP + DP      &20.8G&8.4s&40.6G&13.3s \\ \cline{2-6}
                           & ZeroPP  &12.1G&4.1s&23.8G&9.2s\\ \hline
\end{tabular}
\end{small}
\end{center}
\end{table}

We first demonstrate the performance gap between ZeroPP and 3D parallelism with fixed parallelism degree. In Table \ref{table:main_exp}, we present the iteration time and the maximum allocated GPU memory under different settings. The experimental results align with our theoretical expectations, yet there are notable observations that require further elucidation.

The underperformance of ZB-V can be primarily attributed to two aspects. Firstly, though ZB-B should maintain fixed GPU memory consumption based on the theoretical analysis, the open-source codebase that the paper~\cite{qi2023zero} presented exhibits an increase in memory consumption as the batch size grows in the testings. Secondly, the ZB-V implementation lacks compatibility with numerous 3D parallel optimization techniques, resulting in suboptimal performance.

For ZeroPP, we employ recomputation to reduce the memory footprint of activations. The throughput of ZeroPP with recomputation significantly surpasses that of 3D parallelism.

\subsection{Comparison to 3D parallelism under Equivalent Memory Constraints}
\begin{figure}[ht]
\begin{center}
\centerline{\includegraphics[width=0.7\linewidth]{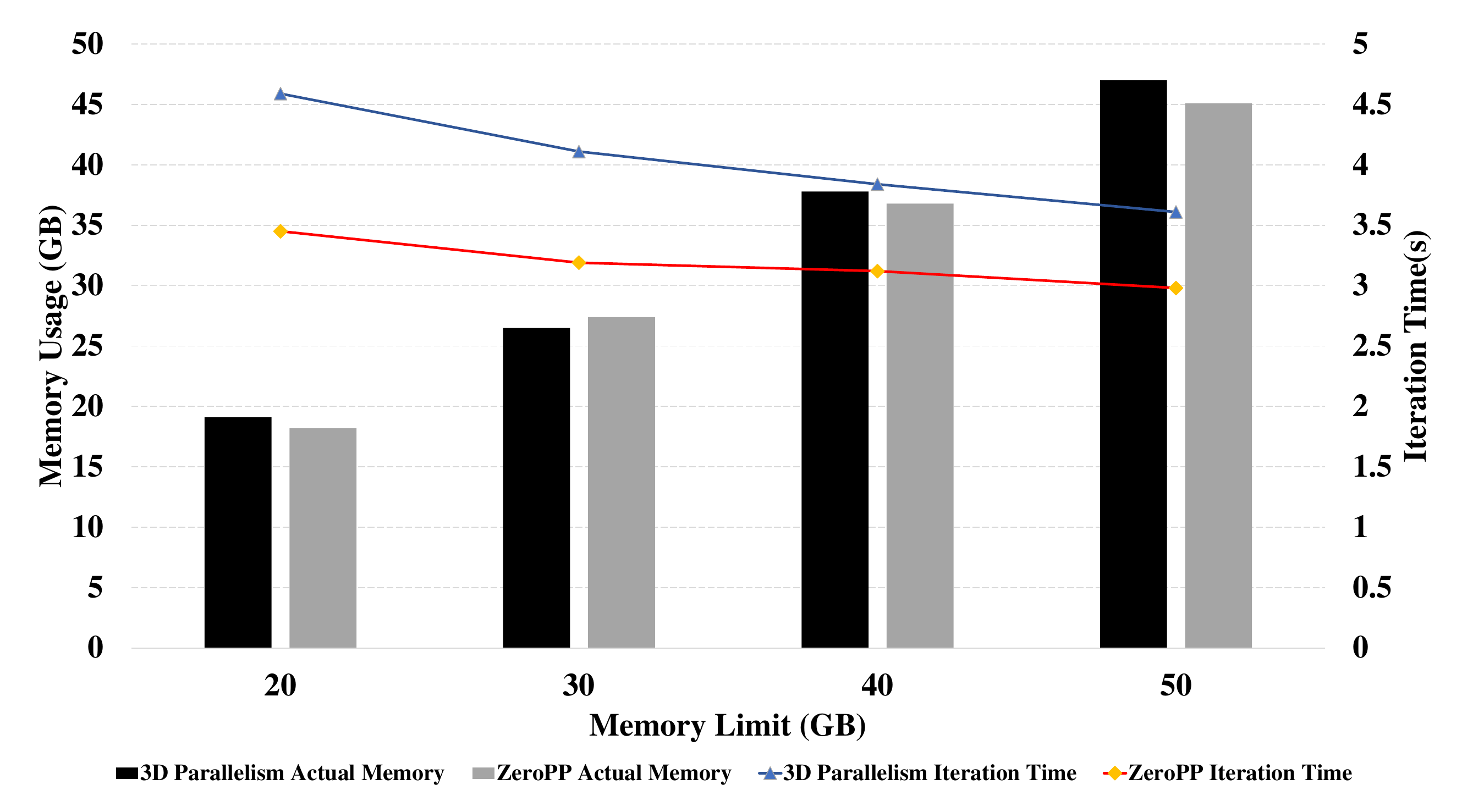}}
\caption{
Comparison of the iteration time and actual GPU memory consumption between ZeroPP and 3D parallelism under equivalent GPU memory constraints.}
\label{fig:memory}
\end{center}
\vskip -0.2in
\end{figure}

In order to provide a clearer comparison of the performance of 3D parallelism and ZeroPP in practical applications, we evaluate the maximum throughput of the two methods under equivalent GPU memory constraints. The experimental results are presented in Figure \ref{fig:memory}.

We conduct experiments using a 14.6B model on 4-node A800 machines with 32 GPUs and explore the hyperparameter space of ZeroPP and 3D parallelism to find the optimal configurations under the same GPU memory limitations. The PP size is uniformly set to 4 and the global batch size is fixed to 192. Due to resource constraints, ZeroPP does not perform DP across nodes, whereas 3D parallelism adjusts the DP and TP sizes within each node. For ZeroPP, we search for the scheduling unit size, the number of stages on each device, and whether to utilize recomputation. For 3D parallelism, we explore the TP/DP size and whether to use sequence parallelism.

The experimental results demonstrate that under equivalent memory constraints, ZeroPP exhibits significantly shorter iteration times compared to 3D parallelism. Moreover, as the memory limitation becomes more stringent, the advantage of ZeroPP becomes increasingly evident. This highlights the effectiveness of ZeroPP and its ability to achieve high performance with reduced memory footprint.

\section{Conclusion}
In conclusion, we propose a scalable TP-free distributed framework that leverages PP and ZeRO-3 efficiently. A ZeRO-compatible PP schedule is proposed which combines ZeRO-3 with high communication efficiency and reduced memory usage. By decoupling gradient computations, near-zero bubble pipeline parallelism is achieved. An efficient recomputation scheme is also proposed to reduce activation memory consumption in ZeroPP. Furthermore, we also perform theoretical analysis on ZeroPP and prove its efficiency over conventional 3D parallelism. We look forward to further applying and optimizing ZeroPP to expansive large-scale training scenes.

However, ZeroPP encounters challenges pertaining to the activation memory, and memory consumption is suboptimal when recomputation is not included. Nonetheless, we show that our recomputation strategy can efficiently resolve this problem and believe that further optimization such as offloading can be achieved by synergistically combining PP with ZeRO-3.

\printbibliography 

@inproceedings{kenton2019bert,
  title={Bert: Pre-training of deep bidirectional transformers for language understanding},
  author={Kenton, Jacob Devlin Ming-Wei Chang and Toutanova, Lee Kristina},
  booktitle={Proceedings of naacL-HLT},
  volume={1},
  pages={2},
  year={2019}
}

@article{brown2020language,
  title={Language models are few-shot learners},
  author={Brown, Tom and Mann, Benjamin and Ryder, Nick and Subbiah, Melanie and Kaplan, Jared D and Dhariwal, Prafulla and Neelakantan, Arvind and Shyam, Pranav and Sastry, Girish and Askell, Amanda and others},
  journal={Advances in neural information processing systems},
  volume={33},
  pages={1877--1901},
  year={2020}
}

@article{touvron2023llama,
  title={Llama: Open and efficient foundation language models},
  author={Touvron, Hugo and Lavril, Thibaut and Izacard, Gautier and Martinet, Xavier and Lachaux, Marie-Anne and Lacroix, Timoth{\'e}e and Rozi{\`e}re, Baptiste and Goyal, Naman and Hambro, Eric and Azhar, Faisal and others},
  journal={arXiv preprint arXiv:2302.13971},
  year={2023}
}

@article{dosovitskiy2020image,
  title={An image is worth 16x16 words: Transformers for image recognition at scale},
  author={Dosovitskiy, Alexey and Beyer, Lucas and Kolesnikov, Alexander and Weissenborn, Dirk and Zhai, Xiaohua and Unterthiner, Thomas and Dehghani, Mostafa and Minderer, Matthias and Heigold, Georg and Gelly, Sylvain and others},
  journal={arXiv preprint arXiv:2010.11929},
  year={2020}
}

@misc{du2021all,
  title={All nlp tasks are generation tasks: A general pretraining framework},
  author={Du, Zhengxiao and Qian, Yujie and Liu, Xiao and Ding, Ming and Qiu, Jiezhong and Yang, Zhilin and Tang, Jie},
  journal={arXiv preprint arXiv:2103.10360},
  volume={18},
  year={2021},
  publisher={Mar}
}

@article{huang2019gpipe,
  title={Gpipe: Efficient training of giant neural networks using pipeline parallelism},
  author={Huang, Yanping and Cheng, Youlong and Bapna, Ankur and Firat, Orhan and Chen, Dehao and Chen, Mia and Lee, HyoukJoong and Ngiam, Jiquan and Le, Quoc V and Wu, Yonghui and others},
  journal={Advances in neural information processing systems},
  volume={32},
  year={2019}
}

@inproceedings{narayanan2019pipedream,
  title={PipeDream: Generalized pipeline parallelism for DNN training},
  author={Narayanan, Deepak and Harlap, Aaron and Phanishayee, Amar and Seshadri, Vivek and Devanur, Nikhil R and Ganger, Gregory R and Gibbons, Phillip B and Zaharia, Matei},
  booktitle={Proceedings of the 27th ACM Symposium on Operating Systems Principles},
  pages={1--15},
  year={2019}
}

@inproceedings{narayanan2021efficient,
  title={Efficient large-scale language model training on gpu clusters using megatron-lm},
  author={Narayanan, Deepak and Shoeybi, Mohammad and Casper, Jared and LeGresley, Patrick and Patwary, Mostofa and Korthikanti, Vijay and Vainbrand, Dmitri and Kashinkunti, Prethvi and Bernauer, Julie and Catanzaro, Bryan and others},
  booktitle={Proceedings of the International Conference for High Performance Computing, Networking, Storage and Analysis},
  pages={1--15},
  year={2021}
}

@article{korthikanti2023reducing,
  title={Reducing activation recomputation in large transformer models},
  author={Korthikanti, Vijay Anand and Casper, Jared and Lym, Sangkug and McAfee, Lawrence and Andersch, Michael and Shoeybi, Mohammad and Catanzaro, Bryan},
  journal={Proceedings of Machine Learning and Systems},
  volume={5},
  year={2023}
}

@article{vaswani2017attention,
  title={Attention is all you need},
  author={Vaswani, Ashish and Shazeer, Noam and Parmar, Niki and Uszkoreit, Jakob and Jones, Llion and Gomez, Aidan N and Kaiser, {\L}ukasz and Polosukhin, Illia},
  journal={Advances in neural information processing systems},
  volume={30},
  year={2017}
}

@article{chen2016training,
  title={Training deep nets with sublinear memory cost},
  author={Chen, Tianqi and Xu, Bing and Zhang, Chiyuan and Guestrin, Carlos},
  journal={arXiv preprint arXiv:1604.06174},
  year={2016}
}

@article{lamy2023breadth,
  title={Breadth-First Pipeline Parallelism},
  author={Lamy-Poirier, Joel},
  journal={Proceedings of Machine Learning and Systems},
  volume={5},
  year={2023}
}

@inproceedings{rajbhandari2020zero,
  title={Zero: Memory optimizations toward training trillion parameter models},
  author={Rajbhandari, Samyam and Rasley, Jeff and Ruwase, Olatunji and He, Yuxiong},
  booktitle={SC20: International Conference for High Performance Computing, Networking, Storage and Analysis},
  pages={1--16},
  year={2020},
  organization={IEEE}
}

@article{2021Colossal,
  title={Colossal-AI: A Unified Deep Learning System For Large-Scale Parallel Training},
  author={ Li, Shenggui  and  Fang, Jiarui  and  Bian, Zhengda  and  Liu, Hongxin  and  Liu, Yuliang  and  Huang, Haichen  and  Wang, Boxiang  and  You, Yang },
  year={2021},
}

@misc{ddp,
  author = {Shen Li},
  year = {2024},
  url = {https://pytorch.org/tutorials/intermediate/ddp_tutorial.html},
  urldate = {2024},
  title = {Getting Started with Distributed Data Parallel}
}

@inproceedings{oh2022out,
  title={Out-of-order backprop: An effective scheduling technique for deep learning},
  author={Oh, Hyungjun and Lee, Junyeol and Kim, Hyeongju and Seo, Jiwon},
  booktitle={Proceedings of the Seventeenth European Conference on Computer Systems},
  pages={435--452},
  year={2022}
}

@article{paszke2019pytorch,
  title={Pytorch: An imperative style, high-performance deep learning library},
  author={Paszke, Adam and Gross, Sam and Massa, Francisco and Lerer, Adam and Bradbury, James and Chanan, Gregory and Killeen, Trevor and Lin, Zeming and Gimelshein, Natalia and Antiga, Luca and others},
  journal={Advances in neural information processing systems},
  volume={32},
  year={2019}
}

@article{shoeybi2019megatron,
  title={Megatron-lm: Training multi-billion parameter language models using model parallelism},
  author={Shoeybi, Mohammad and Patwary, Mostofa and Puri, Raul and LeGresley, Patrick and Casper, Jared and Catanzaro, Bryan},
  journal={arXiv preprint arXiv:1909.08053},
  year={2019}
}

@inproceedings{wang2022tesseract,
  title={Tesseract: Parallelize the tensor parallelism efficiently},
  author={Wang, Boxiang and Xu, Qifan and Bian, Zhengda and You, Yang},
  booktitle={Proceedings of the 51st International Conference on Parallel Processing},
  pages={1--11},
  year={2022}
}

@inproceedings{xu2023efficient,
  title={An efficient 2d method for training super-large deep learning models},
  author={Xu, Qifan and You, Yang},
  booktitle={2023 IEEE International Parallel and Distributed Processing Symposium (IPDPS)},
  pages={222--232},
  year={2023},
  organization={IEEE}
}

@article{wang20212,
  title={2.5-dimensional distributed model training},
  author={Wang, Boxiang and Xu, Qifan and Bian, Zhengda and You, Yang},
  journal={arXiv e-prints},
  pages={arXiv--2105},
  year={2021}
}

@article{bian2021maximizing,
  title={Maximizing parallelism in distributed training for huge neural networks},
  author={Bian, Zhengda and Xu, Qifan and Wang, Boxiang and You, Yang},
  journal={arXiv preprint arXiv:2105.14450},
  year={2021}
}

@article{qi2023zero,
  title={Zero Bubble Pipeline Parallelism},
  author={Qi, Penghui and Wan, Xinyi and Huang, Guangxing and Lin, Min},
  journal={arXiv preprint arXiv:2401.10241},
  year={2023}
}

@article{dean2012large,
  title={Large scale distributed deep networks},
  author={Dean, Jeffrey and Corrado, Greg and Monga, Rajat and Chen, Kai and Devin, Matthieu and Mao, Mark and Ranzato, Marc'aurelio and Senior, Andrew and Tucker, Paul and Yang, Ke and others},
  journal={Advances in neural information processing systems},
  volume={25},
  year={2012}
}

@article{wang2023zero++,
  title={Zero++: Extremely efficient collective communication for giant model training},
  author={Wang, Guanhua and Qin, Heyang and Jacobs, Sam Ade and Holmes, Connor and Rajbhandari, Samyam and Ruwase, Olatunji and Yan, Feng and Yang, Lei and He, Yuxiong},
  journal={arXiv preprint arXiv:2306.10209},
  year={2023}
}

@article{black2022gpt,
  title={Gpt-neox-20b: An open-source autoregressive language model},
  author={Black, Sid and Biderman, Stella and Hallahan, Eric and Anthony, Quentin and Gao, Leo and Golding, Laurence and He, Horace and Leahy, Connor and McDonell, Kyle and Phang, Jason and others},
  journal={arXiv preprint arXiv:2204.06745},
  year={2022}
}

@article{smith2022using,
  title={Using deepspeed and megatron to train megatron-turing nlg 530b, a large-scale generative language model},
  author={Smith, Shaden and Patwary, Mostofa and Norick, Brandon and LeGresley, Patrick and Rajbhandari, Samyam and Casper, Jared and Liu, Zhun and Prabhumoye, Shrimai and Zerveas, George and Korthikanti, Vijay and others},
  journal={arXiv preprint arXiv:2201.11990},
  year={2022}
}

@inproceedings{rajbhandari2021zero,
  title={Zero-infinity: Breaking the gpu memory wall for extreme scale deep learning},
  author={Rajbhandari, Samyam and Ruwase, Olatunji and Rasley, Jeff and Smith, Shaden and He, Yuxiong},
  booktitle={Proceedings of the international conference for high performance computing, networking, storage and analysis},
  pages={1--14},
  year={2021}
}

@article{xu2020megatron,
  title={MEGATRON-CNTRL: Controllable story generation with external knowledge using large-scale language models},
  author={Xu, Peng and Patwary, Mostofa and Shoeybi, Mohammad and Puri, Raul and Fung, Pascale and Anandkumar, Anima and Catanzaro, Bryan},
  journal={arXiv preprint arXiv:2010.00840},
  year={2020}
}

\end{document}